\documentclass[runningheads]{llncs}
\usepackage[american, english, USenglish]{babel}
\usepackage{enumitem}
\usepackage{graphicx}
\usepackage{todonotes}
\usepackage{xspace}
\usepackage{cleveref}
\usepackage{tikz}
\usepackage{listings, listings-rust}
\usepackage[utf8]{inputenc}
\usepackage{lstautogobble}
\usepackage{pgfgantt}
\usepackage{subcaption}
\usepackage{wrapfig}
\usepackage{hyperref}
\usepackage[hyphenbreaks]{breakurl}
\usetikzlibrary{automata, positioning, arrows}
\newcommand*\rtlola{\textsc{RTLola}\xspace}

\definecolor{bluekeywords}{rgb}{0.13, 0.13, 1}
\definecolor{greentypes}{rgb}{0, 0.5, 0}
\definecolor{orangecomments}{rgb}{1, 0.5, 0.1}
\definecolor{redstrings}{RGB}{171, 114, 2}
\definecolor{graynumbers}{rgb}{0.5, 0.5, 0.5}
\definecolor{goldcomments}{rgb}{0.6, 0.4, 0.08}

\newcommand{\sOutput}[0]{\texttt{Event}\xspace}
\newcommand{\sOutputs}[0]{\texttt{Events}\xspace}
\newcommand{\sInput}[0]{\texttt{Verdict}\xspace}

\newcommand{\monInput}[0]{\texttt{Monitor Input}\xspace}
\newcommand{\monOutput}[0]{\texttt{Monitor Output}\xspace}
\newcommand{\eventConversion}[0]{\texttt{Event Conversion}\xspace}
\newcommand{\verdictConversion}[0]{\texttt{Verdict Conversion}\xspace}

\lstdefinelanguage{Lola}{
  keywords=[0]{input, output, trigger, constant, import, spawn, eval, close, with, when},
  keywordstyle=[0]\bfseries\color{bluekeywords},
  keywords=[1]{if, then, else, aggregate, defaults, offset, by, or, to, sin, cos, abs, hold, over, using},
  keywords=[2]{Variable, String, Int, Int8, Int64, UInt, UInt8, UInt64, Bool, Float32, Float64, Float, @1Hz, @5Hz, @10Hz, @100mHz, @1kHz, @1min},
  keywordstyle=[2]\color{greentypes},
    sensitive=false,
    comment=[l]{//},
    morecomment=[s]{/*}{*/},
    morestring=[b]',
    morestring=[b]"
}
\title{Monitoring Unmanned Aircraft: Specification, Integration, and Lessons-learned\thanks{This work was partially supported by the Aviation Research Program LuFo of the German Federal Ministry for Economic Affairs and Energy as part of "Volocopter Sicherheitstechnologie zur robusten eVTOL Flugzustandsabsicherung durch formales Monitoring"(No. 20Q1963C).
}}

\author{Jan Baumeister\inst{1} \and Bernd Finkbeiner\inst{1} \and Florian Kohn\inst{1} \and Florian L\"ohr\inst{2} \and Guido Manfredi\inst{2} \and Sebastian Schirmer\inst{3} \and Christoph Torens\inst{3}}
\authorrunning{Baumeister et al.}
\titlerunning{Monitoring Unmanned Aircraft}

\institute{
CISPA Helmholtz Center for Information Security \\
\and Volocopter GmbH \and German Aerospace Center (DLR)
}

\renewcommand{\todo}[1]{}
\renewcommand{\lstinline}[1]{\texttt{#1}}

\begin{document}
\maketitle
\begin{abstract}
This paper reports on the integration of runtime monitoring into fully-electric aircraft designed by Volocopter, a German aircraft manufacturer of electric multi-rotor helicopters. The runtime monitor recognizes hazardous situations and system faults. Since the correct operation of the monitor is critical for the safety of the aircraft, the development of the monitor must follow strict aeronautical standards. This includes the integration of the monitor into different development environments, such as log-file analysis, hardware/software-in-the-loop testing, and test flights.
We have used the stream-based monitoring framework RTLola to generate monitors for a range of requirements.
In this paper, we present representative monitoring specifications and our lessons learned from integrating the generated monitors.  
Our main finding is that the specification and the integration need to be decoupled, because the specification remains stable throughout the development process, whereas the different development stages require a separate integration of the monitor into each environment. We achieve this decoupling with a novel abstraction layer in the monitoring framework that adapts the monitor to each environment without affecting the core component generated from the specification. The decoupling of the integration has also allowed us to react quickly to the frequent changes in the hardware and software environment of the monitor due to the fast-paced development of the aircraft in a startup company.
\end{abstract}
\todo{Into UAV}
\section{Introduction}

The new generation of fully-electric aircraft pioneered by companies
like Volocopter promises a revolution in urban air
mobility. Fully-electric aircraft air taxis, cargo drones, and
longer-range passenger aircraft will provide transit solutions that
are emission-free and thus more sustainable and efficient than
traditional forms of air transport. A critical part of the safety
engineering of such aircraft is to analyze log-files and tests, as well as 
the real-time data obtained during the actual flight, so that
the health status of the system can be assessed and mitigation
procedures can be initiated when needed.  In this paper, we report on
the design and integration of formally specified monitors into
aircraft developed by Volocopter, based on the monitoring framework
RTLola. The goal of our collaboration over the past three years has
been to explore the benefits and challenges of applying formal runtime
verification within the strict aeronautical standards of aircraft 
development.

Volocopter specializes in the design, manufacturing, and operations of
electric Vertical Takeoff and Landing (eVTOL) vehicles.  The company
targets Urban Air Mobility (UAM) operations, i.e., passenger and cargo
transportation above and around cities. These operations involve high
population density on the ground and high traffic density in the air.
Consequently, all developments must meet the highest level of safety
similar to airliners: one failure for every billion hours flown.
To ensure such a level of safety, the design of the vehicles follows
aeronautical standards, especially SAE's ARP4754b \cite{ARP4754B} to
ensure the coherency between the concept of operation, requirements, design, and
implementation. The development cycle described in this
standard uses a layered approach with multiple verification and
validation steps.

RTLola~\cite{streamlab,DBLP:journals/corr/abs-2003-12477} is a formal monitoring framework that consists of a
stream-based specification language for real-time properties, an
interpreter, and compilers into software- and hardware-based execution
platforms.  An RTLola specification of hazardous situations and system
failures is statically analyzed in terms of consistency and resource
usage and then automatically translated into an FPGA-based
monitor. This approach leads to highly efficient, parallelized monitors
with formal guarantees on the noninterference of the monitor with the
normal operation of the monitored system.

Previous case studies with RTLola~\cite{DBLP:conf/cav/BaumeisterFSST20} and similar frameworks, such as R2U2~\cite{10.1007/978-3-031-37709-9_23}
and Copilot~\cite{perez2020copilot}, have already shown that properties that are critical for
the safety of the aircraft can readily be expressed in such formal
languages and that the resulting monitors can be integrated into real
systems. Our ambition has been to go beyond such one-time
applications, and integrate the specified monitors into the complete
development process. This means that the generated monitors are not
only integrated into the specific setup of the case study, but rather
are continuously adapted according to the needs of the development
process.

We consider monitoring in all stages of the development process. Initially, the
role of the monitor is to annotate log-files and guide the user during
an offline analysis. Next, the monitor validates data from
test-benches that check that external components conform to their
specifications, such as delivering data within deadlines. Finally, the
monitor validates safety requirements during test flights. The
monitoring specifications are based on the requirements of the various
regulatory authorities and cover a range of safety-critical
requirements from single-component checks to system-level health.

Our main finding is that the specification and the integration need to
be decoupled, because the specification remains stable throughout the
development process, whereas the different development stages require
a separate integration of the monitor into each environment. We
achieve this decoupling with a novel abstraction layer in the
monitoring framework that adapts the monitor to each environment
without affecting the core component generated from the specification.
In the abstraction layer, the monitor is framed with two new
components, the \emph{event conversion} and the \emph{verdict converison}.  The decoupling of the integration has also allowed us to
react quickly to the frequent changes in the hardware and software
environment of the monitor due to the fast-paced development of the
aircraft in a startup company.

\subsection{Related Work}
Runtime monitoring is a scalable dynamic verification approach that has been applied to a variety of domains~\cite{junges2021runtime,10.1007/978-3-031-37703-7_17}.
For cyber-physical systems, many monitoring tools exist~\cite{Bartocci2018,perez2020copilot,DBLP:conf/cav/BaumeisterFSST20}, 
but despite integration being an important part of the usage of monitoring ~\cite{falcone2021taxonomy}, tools are often specific to certain environments and leave embedding in different environments to the user, i.e., the user needs to establish a connection, parse received events, and forward it to the monitor.
For some specific environments, these user efforts are reduced.
For instance, SOTER~\cite{DBLP:conf/dsn/DesaiGSST19} a specification language that is based on the P language~\cite{DBLP:conf/pldi/DesaiGJQRZ13}, 
was recently extended~\cite{10.1007/978-3-030-60508-7_10} to produce code for the Robot Operating System (ROS), which allows to just specify which ROS topics are subscribed and published.
Similarly, TeSSLa features keywords to subscribe and publish ROS topics~\cite{10.1007/978-3-030-60508-7_10}.
A more generic approach is pursued by R2U2 Version 3.0~\cite{10.1007/978-3-031-37709-9_23} which allows to specify C-like structs.
This makes it easy for engineers to receive structs and just forward them to the monitor unit.
In this work, we foster this kind of generalization by providing an automatic mapping of the received and the forwarded events. \todo{sebas: can we write something like this?}
\section{Stream-based Monitoring}
\label{sec:rtlola}
\rtlola is a real-time monitoring framework~\cite{streamlab} \todo{sebas: when refering to RTLola, stick to framework/toolkit/.., dont switch} aimed at, but not exclusively applicable to, cyber-physical systems.
At its core is a stream-based specification language that distinguishes between two kinds of streams: Input streams represent sensor readings from the system under observation. Output streams perform computations over these input streams and other output streams.
Special kinds of output streams, called triggers, define violations based on boolean conditions.
Equipped with a message, they notify the system operator when a violation is detected.
Consider the following example:
\begin{lstlisting}
	input altitude: Float
	output average_alt @1Hz	 := altitude.aggregate(over: 60s, using: avg).defaults(to: 0.0) 
	trigger average_alt > 300.0
\end{lstlisting}
In this example, the monitor observes the altitude of the system through the input stream \lstinline{altitude}.
The output stream \lstinline{average\_alt} aggregates all values of this input stream over the last minute and computes the average of these values.
It also highlights the real-time capabilities of \rtlola.
By explicitly annotating the output stream with a frequency, the monitor cannot only react to events but also proactively perform computations.
More concretely, the output stream evaluates at a fixed frequency of \lstinline{1Hz}.
The final defined trigger then notifies an operator if the average altitude is above $300$.
\section{Setup}
\label{sec:setup}
All components that are integrated into aircrafts designed by Volocopter need to
follow aeronautical standards, especially SAE's ARP4754b \cite{ARP4754B}.
This standard ensures that the concept of operation, requirements, design, and implementation are coherent.
In general, it describes a development cycle using a layered approach with multiple verification and validation steps,
i.e., new components are validated in different environments that get closer to the operation with each step.
The monitor can provide valuable feedback when new components undergo the aforementioned validation steps.
This feedback includes statistical assessments or violations of given requirements.
Yet, the monitor as a safety-critical component needs to be evaluated in the same manner.
\subsection{Monitoring Applications}
\label{sec:mon-app}
This section presents four applications that highlight the benefit of the monitors feedback during the development of new components:
\begin{enumerate}
	\item \emph{Debugging}
	      A monitor is developed alongside the component giving full information about its internal state.
	      During execution of the system, the monitor checks whether the component works as intended by the developer.
	\item \emph{Validation}
	      The monitor is developed independently of the component and checks its behavior based on the inputs and outputs of defined test cases.
	      The monitor output on these test cases is then used as a report for internal validation, validation of components by external companies, or as proof of conformity for aviation authorities.
	\item \emph{Pre-Post-Flight Analysis}
	      Before the flight, the monitor checks whether all necessary components are operational.
	      After the flight, the monitor computes more sophisticated information to better evaluate the flight and detect irregularities that were not detected during the flight.
	\item \emph{In-Flight Analysis / Safe Integration}
	      The monitor communicates with the remote operator. e.g., through the User Interface of the ground control station, to provide feedback about the safety of the drone.
	      It validates the correctness of individual components to ensure a safe flight or monitors the flight operation.
	      For instance, it supports the pilot by checking that the drone stays within safe flight parameters such as a geofence.
\end{enumerate}
Before presenting concrete specifications for each application in \Cref{sec:case_study}, we elaborate on the validation of a monitor.
 
\subsection{Development Cycle for the Monitor}
\label{sec:dev-cycle}
This section introduces the four environments into which the monitor must be integrated to validate its correctness.
\begin{enumerate}
	\item \emph{Log-File Analysis} This step evaluates the functional correctness of the specification.
	      We test the generated monitors against traces that violate or satisfy the specification and analyze the output of the monitor.
	\item \emph{Software-in-the-Loop (SiL)} The monitor interacts with simulated systems and environments.
	      This step is crucial for a runtime monitor since most temporal behaviors are not visible until these tests.
	\item \emph{Hardware-in-the-loop (HiL)} This step is similar to the SiL environment. However, the monitor and the system run on the actual resource-constrained hardware used in the aircraft.
	      This setup brings even more time-related effects to the evaluation and allows an evaluation with replayed flight data.
	\item \emph{Flight Testing}
	     Running the monitor in parallel with the flying aircraft allows for assessing the impact of all effects coming from the aircraft, the ground system, and the environment.
\end{enumerate}
The integration of the monitors in the different validation environments poses new challenges for the monitoring framework.
In our experience, each step in the development process relies on different ways of communication.
For instance, in the log-file analysis, events are processed in CSV-format, while during test flights, the communication with the monitor uses a custom protocol over TCP.
Yet, the changes in the monitor should be as minimal as possible to simplify its validation.
Specifically, the specification has to remain unchanged after the \emph{Log-file Analysis} as otherwise its functional correctness is not guaranteed anymore.
\section{Abstract Integration}
\label{sec:integration}
\tikzset{
	->,
	>=stealth', 
	node distance=3cm, 
	every state/.style={thick, fill=gray!10, shape=rectangle, align=center, minimum width=1.6cm}, 
	initial text=$$, 
	on grid,
	minimum width=0cm,
	minimum height=0cm
}
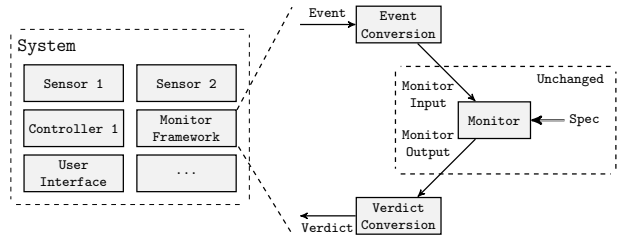
\begin{figure}[t]
	\centering
	\scalebox{0.6}{
		\begin{tikzpicture}
			\node[state] (mon) {\texttt{Monitor}};
			\node[state, below left of=mon] (sink) {\texttt{Verdict}\\\texttt{Conversion}};
			\node[state, above left of=mon] (src) {\texttt{Event}\\\texttt{Conversion}};
			\node[left of=src, node distance=2.3cm] (in) {};
			\node[left of=sink, node distance=2.3cm] (out) {};

			\node[right of=mon, node distance=2cm] (spec) {\texttt{Spec}};

			\node[above left=3.1cm of mon, yshift=-1cm] (leftAbove) {};
			\node[below left=3.1cm of mon, yshift=1cm] (leftBelow) {};
			\node[below right=3.1cm of mon, yshift=1cm, xshift=0.5cm] (rightBelow) {};
			\node[above right=3.1cm of mon, yshift=-1cm, xshift=0.5cm] (rightAbove) {};
			\node[above right=3.1cm of mon, yshift=-1.3cm, xshift=-0.5cm] (desc) {\texttt{Unchanged}};

			\draw[thick]
			(in) edge[above] node[align = center, minimum size=0.5cm] {\sOutput} (src)
			(src) edge[below] node[align = center, minimum size=1cm, xshift=-0.4cm]{\texttt{Monitor} \\ \texttt{Input}} (mon)
			(mon) edge[above] node[align = center, minimum size=1cm, xshift=-0.4cm] {\texttt{Monitor} \\ \texttt{Output}} (sink)
			(sink) edge[below] node[align = center, minimum size=0.5cm]{\sInput} (out);

			\draw[-, dashed, thick] (rightAbove) -- (leftAbove) -- (leftBelow) -- (rightBelow) -- (rightAbove);
			\draw[double] (spec) -- (mon);

			\node [state, left=4.7cm of src, minimum width=2.2cm, yshift=-1.3cm] (s2) {\texttt{Sensor 2}};
			\node [state, left=2.5cm of s2, minimum width=2.2cm] (s1) {\texttt{Sensor 1}};
			\node [state, below=1cm of s1, minimum width=2.2cm] (c1) {\texttt{Controller 1}};
			\node [state, below=1cm of s2, minimum width=2.2cm] (monf) {\texttt{Monitor}\\ \texttt{Framework}};
			\node [below=1cm of s2, minimum width=2.2cm, minimum height=0.8cm, xshift=0.9cm] (monsr) {};
			\node [state, below=1cm of c1, minimum width=2.2cm] (ui) {\texttt{User}\\ \texttt{Interface}};
			\node [state, below=1cm of monf, minimum width=2.2cm] (dots) {\texttt{\dots}};

			\node[above left=3.1cm of s1, yshift=-1cm, xshift=0.8cm] (leftAboveN) {};
			\node[below left=3.1cm of ui, yshift=1.5cm, xshift=0.8cm] (leftBelowN) {};
			\node[below right=3.1cm of dots, yshift=1.5cm, xshift=-0.8cm] (rightBelowN) {};
			\node[above right=3.1cm of s2, yshift=-1cm, xshift=-0.8cm] (rightAboveN) {};
			\node[above left=3.1cm of s1, yshift=-1.4cm, xshift=1.6cm] (desc) {\large \texttt{System}};
			\node[above=0.5cm of in] (inN) {};
			\node[below=0.5cm of out] (outN) {};

			\draw[-, dashed] (rightAboveN) -- (leftAboveN) -- (leftBelowN) -- (rightBelowN) -- (rightAboveN);

			\draw[-, dashed, thick] (monsr) -- (outN) (monsr) -- (inN);
		\end{tikzpicture}
	}
	\caption{Overview of the Generalization}
	\label{inte:overview}
\end{figure}
In this section, we present our approach integrating the \rtlola framework~\cite{streamlab} into the different environments described in \Cref{sec:setup}.

\Cref{inte:overview} shows an overview of this approach.
The system on the left side represents the UAV under development.
From a monitoring perspective, the current step in the development cycle does not influence the underlying monitor, only its integration into the system.
This is depicted on the right side of \Cref{inte:overview}.
The monitor framework receives or requests incoming data from the different components of the system (\sOutput), analyses this data, and produces an output (\sInput).
In the center of the monitoring framework is a fixed monitor generated from a formal specification.
This monitor has a fixed representation of the \monInput\ and \monOutput\ that are independent of the integration.

To bridge the communication gap between the system and the monitor, we propose an abstraction layer that translates system outputs to monitor inputs and vice versa.
This abstraction also generalizes the monitor's interface such that no expert knowledge about the concrete monitor is necessary to integrate the monitor.
This abstraction layer can vary depending on the specific integration into the development cycle but allows the monitor to remain unchanged following the idea of decoupling the specification from the integration.

Our approach introduces two translation components: the \eventConversion and the \verdictConversion.
Each component is again split into two parts: \emph{data-acquisition/data-dispatch} and \emph{data-conversion} to allow a generic implementation of the \emph{data-acquisition}/\emph{data-dispatch}.
This results in reusable and maintainable implementation while keeping the changes between the development stages minimal.

In the following, we elaborate on the data-acquisition and data-conversion of the \emph{Event Conversion} in more detail.
The results are transferable to the \emph{Verdict Conversion}.

\subsection{Common Interfaces}
The \eventConversion~is a generic translation layer between the systems output and the monitor input.
During instantiation it validates the mapping between the system output and the monitor input, representing the input streams in the specification.
Hence, it checks that the \sOutputs are a superset of the \monInput avoiding any runtime errors resulting from an invalid mapping.
The \emph{data-acquisition} part of the \eventConversion is handled through the \emph{Event Source} interface, while the \emph{data-conversion} is handled by an \emph{Event Factory}.
Both interfaces are defined in \Cref{fig:event_conversion}.
\begin{figure}[t]
	\begin{subfigure}{0.49\textwidth}
		\begin{lstlisting}[language=Rust, basicstyle=\ttfamily\scriptsize]
		pub trait EventSource {
		  fn next_event(&mut self)
		    -> Result<Event, Error>;
		}
		\end{lstlisting}
		\caption{Event Source Interface}
		\label{fig:eventSource}
	\end{subfigure}
	\begin{subfigure}{0.49\textwidth}
		\begin{lstlisting}[language=Rust, basicstyle=\ttfamily\scriptsize]
		pub trait EventFactory {
		  fn new(map: &InputStreams, cfg: Config)
		    -> Result<Self, Error>;
		  fn create(&mut self, ev: Event)
		    -> Result<MonitorInput, Error>;
		}
		\end{lstlisting}
		\caption{Event Factory Interface}
		\label{fig:eventFactory}
	\end{subfigure}
	\caption{Common interfaces for the Event Conversion.}
	\label{fig:event_conversion}
\end{figure}

The \emph{Event Source} consists of a single function called \texttt{next\_event}.
It is used to communicate to the system that the monitor is ready to accept the next event.
The \emph{Event Factory} as a counterpart has two functions:
The \texttt{new} function gets a description of the input streams derived from the specification and the configuration of the \emph{Event Source}.
It then checks if each input in the specification can be matched with the data provided by the \texttt{Event Factory} implementation.
If successful, it computes a static mapping for each input stream to a data segment in an incoming \sOutput.
The second function \texttt{create} is called for every \sOutput and creates the internal event structure \monInput, given the input mapping.
\paragraph{Implementation.}
We implemented the approach from this section in the \rtlola framework and were able to provide implementations for a variety of \emph{Event Sources} that are independent of the data format they receive.
These include basic file-based input methods such as reading from stdin or a local file, up to network protocols that receive data over UDP, TCP, or MQTT.
We also provide ready-to-use \emph{Event Factories} to parse, for example, data in CSV or PCAP format as well as a binary data parser derived from a user-provided configuration.
\begin{wrapfigure}{r}{0.3\textwidth}
	\vspace{-10pt}
	\begin{lstlisting}[language=Rust, basicstyle=\ttfamily\scriptsize]
		#[derive(ValueFactory)]
		#[factory(prefix)]
		struct GPS {
			lat: Float64,
			lon: Float64
		}
	\end{lstlisting}
	\caption{Interfacing a custom data structure.}
	\label{fig:example:ef}
	\vspace{-10pt}
\end{wrapfigure}
Yet, implementing a custom \emph{Event Factory} still requires knowledge about the structure of the \monInput that is undesired for a successful integration in real-production where implementations need to be maintained by non-monitoring experts.
In the \rtlola framework, we provide further abstractions over the interfaces presented in \Cref{fig:event_conversion} to reduce the required knowledge about the monitoring framework.
These abstractions range from helper implementations encapsulating common functionality to procedural macros that automatically generate implementations of these interfaces.
\Cref{fig:example:ef} shows an example of the macro application.
It demonstrates a simplified version of a GPS-Package, exposing the fields of the struct to input streams named \texttt{GPS\_lat} and \texttt{GPS\_lon}.
\section{Concrete Integration of Representative Specifications}
\label{sec:case_study}
\newcommand\textganttbar[4]{%
  \ganttbar{#1}{#3}{#4}
  \ganttbar[inline,bar label font=\footnotesize]
  {#2}{#3}{#4}
}
This section provides a set of representative specifications to validate our approach presented in \Cref{sec:integration}.
The specifications have been obtained by collaborating with flight engineers or from official RTCA\cite{RTCA} standards and cover all monitoring applications from \Cref{sec:mon-app}.
\Cref{fig:case-study} provides an overview of the concrete specifications and the integration of the generated monitors.

The x-axis in this graph maps each specification to the environments in which the monitor was integrated.
Not all monitors could be validated up to a flight test, but we validated our approach in at least two environments presented in \Cref{sec:dev-cycle}.
In our experience, the provided \emph{Event Sources} are sufficient for all environments. 
For the concrete implementation of the \emph{Event Factories}, we either use the implementation from \Cref{sec:integration} or create new implementations using the macros shown in \Cref{fig:example:ef}.
These implementations do not require any internal knowledge of the monitoring tool as intended by our approach.

In our setup, log-files are usually given in the csv-format, the SiL is a Matlab Simulation or a simplified replay of log-files and the Hil is a concrete replay of the simulated data or flights on the actual hardware.
The test flights were performed on the VoloDrone, a cargo transportation drone that offers highly automated flights with a range of 40km and a payload of up to 200kg.

The y-axis of the graph in \Cref{fig:case-study} maps the specification to the application. We separately validated the specifications with a log-file analysis and refined the requirements before integrating the monitors.
We published the specifications on Github\footnote{\url{https://github.com/reactive-systems/rtlola-uav-specifications}} after replacing some sensitive information, e.g., by replacing some streams with arbitrary constants.
The rest of the section describes the general idea of the specifications and refers to the concrete monitoring application.

\begin{figure}[tbp]
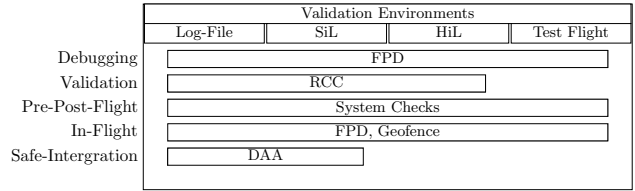

  \centering
  \scalebox{0.65}{
    \begin{ganttchart}[
        y unit title=0.4cm,
        y unit chart=0.5cm,
        title label anchor/.style={below=-1.6ex},
        title left shift=.05,
        title right shift=-.05,
        title height=1,
        progress label text={},
        bar height=0.7,
        group right shift=0,
        group top shift=.6,
        group height=.3
      ]{1}{20}

      \gantttitle{Validation Environments}{20} \\
      \gantttitle{Log-File}{5}
      \gantttitle{SiL}{5}
      \gantttitle{HiL}{5}
      \gantttitle{Test Flight}{5}\\

      \textganttbar{Debugging}{FPD}{2}{19} \\
      \textganttbar{Validation}{RCC}{2}{14} \\
      \textganttbar{Pre-Post-Flight}{System Checks}{2}{19} \\
      \textganttbar{In-Flight}{FPD, Geofence}{2}{19} \\
      \textganttbar{Safe-Intergration}{DAA}{2}{9} \\
    \end{ganttchart}}
  \caption{Overview of the concrete integrations that have been performed in the research project}
  \label{fig:case-study}
\end{figure}
\begin{figure}[t]
  \begin{lstlisting}
		input rpm : Int64
		input src : UInt8
		output rpm_1 eval when src == ROTOR_1 with abs(rpm)
		output rpm_2 eval when src == ROTOR_2 with abs(rpm)
		...
		output rpm_on_check := avg(rpm_1.hold(or: 0), rpm_2.hold(or: 0), ...) > $\epsilon_\mathit{rpm\_on}$
		output rpm_on @1s := rpm_on_check.aggregate(over: 1s, using: avg, or: 0.0) > $\epsilon_\mathit{rpm\_on\_per}$
		...
		output phase_1 := $\neg$take_off $\wedge$ $\neg$landed $\wedge$ rpm_on $\wedge$ $\neg$rpm_in_air
	\end{lstlisting}
  \caption{Excerpt of the specification of the Flight-Phase-Detection}
  \label{fig:spec:fpd}
\end{figure}

\paragraph{Flight-Phase-Detection (FPD)}
The FPD specification detects different flight phases, helping the debugging of correct automated flights.
In the log-file analysis, the monitor annotates previous test flights pointing the engineer to critical points, e.g., when no clear phase could be detected.
In the software and hardware simulation, we evaluate the handling of asynchronous inputs and the timing of the monitor.
For a final flight test, the monitor was integrated into the ground station to check if a flight phase is always detected moving the monitor also to the in-flight application.

\Cref{fig:spec:fpd} presents partially the specification for the FPD.
It gets data from several sensors and computes binary flags describing the current state of the drone.
One example is the \texttt{rpm\_one\_check} flag that compares the average rotations per minute of all rotors against a threshold.
In general, a simple state machine then decides based on these flags if a flight phase is detected and which one.
However, the data of the sensors arrives asynchronously with different frequencies and we need to synchronize the flags for the comparison.
For this synchronization, the streams \texttt{rpm\_on} aggregate over the corresponding flags, computing the percentage of how often the condition is satisfied during the last second.
This value is then used in \texttt{phase\_1} stream for the flight phase detection instead of asynchronous \texttt{rpm\_one\_check} flag.

\begin{figure}[t]
  \begin{lstlisting}
		/// Property 1: Log message increment
		output valid_seq_number := seq_number = seq_number.offset(by: -1, or: -1) + 1
		/// Property 7: RC fallback test
		output main_fallback_valid_dyn
			spawn when lost_connection_to_master
			close when switch_to_secondary $\vee$ both_rc_disconnected
			eval @200ms with false
		output main_fallback_valid @true := main_fallback_valid_dyn.hold(or: true)
		\end{lstlisting}
  \caption{Excerpt of the requirements specific to one RCC}
  \label{fig:sec:rcs:single}
\end{figure}
\paragraph{Remote-Control System (RCS)}
Assuring a safe development is especially challenging when combining in-house products with commercial off-the-shelf hardware or software products.
In our example, we validated the correctness of an RCS that receives flight commands from different sources and dependent on the configuration decides which source should be used by the system.
More concretely, we used \rtlola to validate that the requirements given to the company developing the RCS are satisfied by the resulting product.
Besides the in-house validation, this approach comes with certification evidence that can be submitted to the safety agency for the certification process.

As a redundant system, the RCS runs several instances of remote control computers (RCCs) and unions their output.
\Cref{fig:sec:rcs:single} presents some stream-declarations for requirements validating each RCC individually.
This specification includes checks of simple invariants such as property one that validates if the sequence number increments, but also includes complex real-time properties exemplified with property seven.
The stream declarations for this property implement a watchdog.
It reports a violation in the case that the connection to the main controller is lost and the RCC does not switch from the main to the secondary controller in a time frame of 200ms.

\paragraph{System Checks}
We developed a specification to validate the system parameters of different sensors.
This specification included requirements monitoring the battery level and voltage drops, pre-flight sensor inconsistencies, and accelerations bound.
Due to the sensible information of this specification that requires knowledge of the complete system, we cannot publish this specification.

\paragraph{Geofence}
Defining a geography volume and a contingency volume, where the UAV will operate and can be used to maneuver in case of a problem, is part of a risk assessment required for a flight permit in the specific category~\cite{SORA}.
This risk assessment also requires a runtime validation that the position of the UAV is within these bounds for which we use the monitors generated from an \rtlola specification.
Similar to the FPD, we integrated the monitor in the ground station to communicate with the remote pilot.
We used the geofence specification from previous case-studies with RTLola \cite{DBLP:conf/cav/BaumeisterFSST20,Schirmer2022} describing the intersection between the flight vehicle line and the geofence polygon.
Further, this paper extends the specification to predict a possible breach of the geofence by computing the minimum distance to each polygon line and to approximate the time until that breach.

\paragraph{Detect-And-Avoid (DAA)}
We use the validation of the DAA function as a representative specification for the safe integration monitoring application.
This function is essential for any UAV flying beyond visual line of sight and ensures that the UAV avoids any collision with the surrounding traffic.
One of the most common sensors in commercial aviation is the ADS-B in receiver which can sense all surrounding aircraft equipped with ADS-B out emitters.
However, this sensor is susceptible to attacks by spoofing, so the RTCA standard \cite{RTCA} demands a safe integration \todo{Guido} in which this sensor needs to be supported by a secondary sensor, usually an "active surveillance" sensor.
Instead of merging both signals, it is common practice to use the active surveillance sensor data to check if parts or all of the ADS-B in signal have been compromised.
The challenges for the \rtlola specification are similar to the flight phase detection.
The specification compares data from sensors with different frequencies and validates these frequencies.
Compared to the FPD, the standard assumes in its validation these frequencies, so a comparison with the last values is sufficient instead of aggregating the data.

\section{Conclusion}
This paper presents the results of our research project investigating the use of runtime monitors implemented in the \rtlola framework for the development of unmanned aircraft systems.
We demonstrate the benefits of decoupling the specification and integration when 
the monitor has to undergo the same development as other safety-critical components, in this safety-critical environment.
To keep the changes for the monitor during the development as minimal as possible,
we presented an abstraction for monitoring frameworks.
This abstraction introduces two layers that translate between system outputs and monitoring inputs and vice versa.
We conducted a large case study to validate our approach and presented representative specifications for different monitoring applications derived from aeronautical safety standards and internal requirements from Volocopter.
In a final step, we performed a test flight where the monitor reported its feedback to the ground control station used by the remote pilot.
From a monitoring perspective, this approach can be used to start the development of automatic contingencies triggered by the monitor instead of notifying the pilot.
\bibliographystyle{splncs04}
\bibliography{references.bib}
\end{document}